\documentclass[aps,pra,reprint,superscriptaddress,longbibliography]{revtex4-1}
\usepackage[colorlinks=true,allcolors=blue]{hyperref}
\usepackage{mathtools}
\usepackage{amssymb}
\usepackage{bm}
\usepackage[normalem]{ulem}

\usepackage{lipsum}
\usepackage[pdftex]{graphicx}
\usepackage{microtype}
\usepackage{verbatim}
\usepackage{todonotes}
\presetkeys{todonotes}{inline,backgroundcolor=yellow}{}
\usepackage{subcaption}
\usepackage{soul}
\usepackage{bbm}
\usepackage{tikz}
\usepackage{bbold}
\usepackage{qcircuit}
\usepackage{amsmath, amssymb}
\usepackage{multirow}
\usepackage{array}
\usepackage{stackengine}

\newcommand{\ket}[1]{{| {#1} \rangle}}
\newcommand{\bra}[1]{{\left\langle {#1} \right|}}

\DeclareMathOperator{\Tr}{Tr}

\begin{document}
\title{Quantum recurrences in the kicked top}
\author{Amit Anand}
\affiliation{Institute for Quantum Computing, University of Waterloo, Waterloo, Ontario, Canada N2L 3G1}
\affiliation{Department of Physics and Astronomy, University of Waterloo, Waterloo, Ontario, Canada N2L 3G1}
\author{Jack Davis}
\affiliation{Institute for Quantum Computing, University of Waterloo, Waterloo, Ontario, Canada N2L 3G1}
\affiliation{Department of Physics and Astronomy, University of Waterloo, Waterloo, Ontario, Canada N2L 3G1}

\author{Shohini Ghose}
\affiliation{Institute for Quantum Computing, University of Waterloo, Waterloo, Ontario, Canada N2L 3G1}
\affiliation{Department of Physics and Computer Science, Wilfrid Laurier University, Waterloo, Ontario, Canada N2L 3C5}
\affiliation{Perimeter Institute for Theoretical Physics, 31 Caroline St N, Waterloo, Ontario, Canada N2L 2Y5}

\begin{abstract}

The correspondence principle plays an important role in understanding the emergence of classical chaos from an underlying quantum mechanics.  
Here we present an infinite family of quantum dynamics that never resembles the analogous classical chaotic dynamics irrespective of dimension.  These take the form of stroboscopic unitary evolutions in the quantum kicked top that act as the identity after a finite number of kicks.  Because these state-independent temporal periodicities are present in all dimensions, their existence represents a universal violation of the correspondence principle.  We further discuss the relationship of these periodicities with the quantum kicked rotor, in particular the phenomenon of quantum anti-resonance.
\end{abstract}

\maketitle

\section{Introduction}
\label{introduction}

The quantum-classical correspondence principle broadly states, in its commonly understood form, that the predictions of a dynamically evolving quantum system should reproduce the predictions of a classical system under appropriate circumstances \cite{fortin_2019_the}.  This sometimes takes the form of a particular limit of some set of parameters that characterize the quantum system (i.e.\ large quantum numbers, vanishing Planck action, etc.).  In such situations the transition may be called a \textit{classical limit} of the quantum system \cite{gutzwiller_1990_chaos}. It is well know that classical systems can display chaotic behaviour - broadly defined as exponential sensitivity to initial conditions. Interestingly, in quantum systems that have such a chaotic classical limit, the correspondence principle is not well understood \cite{Casati_Chirikov_Izraelev_Ford_1979,Izrailev_Shepelyanskii_1980,Casati_Guarneri_1984,Kumari_Ghose_vicinity_2018}. Exploring such systems can thus provide insight to the structural differences between quantum and classical dynamics as well as the fundamental origin of chaotic phenomena.

A useful model studied in this context is the quantum kicked top \cite{haake1987}.  This model is a spin-$j$ system subject to a Floquet evolution (i.e.\ a stroboscopic dynamics).  It is of interest because it lives in a finite-dimensional Hilbert space, its dynamics have a well-defined classical limit ($j\rightarrow\infty$) with an easily tunable degree of chaos via its Hamiltonian parameters, and it is experimentally feasible \cite{Chaudhury_2009, Neill_2016, Krithika_NMR_QKT_2019}.  Furthermore, the alternative representation of any spin-$j$ system as a many-body system of indistinguishable qubits has lead to much work on understanding the surprisingly subtle relationship between dynamical entanglement, Hilbert space dimension, and emergent chaos in the kicked top model \cite{Ghose_Sanders_ent_dynamics_2004, Ghose_chaos_ent_dec_2008, Lombardi_ent_2011, Ruebeck_Pattanayak_2017, Madhok_Dogra_Lakshminarayan_2018, Kumari_Ghose_vicinity_2018, Dogra_exactly_2019}.

In this paper, we probe quantum-classical correspondence in the kicked top, and present a startling result. For certain system parameters, classical chaotic behaviour is not recovered no matter how large the value of the spin quantum number, in contradiction to Bohr correspondence. We analytically and numerically show that in all dimensions (all values of the spin $j$), the kicked top displays several state-independent, temporal periodicities/recurrences: three for integer spin values and two for half-integer spin values. Whereas previous work has explored a specific temporal periodicity in the semiclassical limit \cite{Zou_Wang_pseudo_2022}, the general set of recurrences derived in our analysis have not been previously identified.
Because these recurrences are state-independent and generally occur at large chaoticity values, they have no classical analog and so represent a violation of the correspondence principle. Furthermore, our results show that the transition to classical behaviour does not smoothly vary with the size of the system. Our analysis also resolves previous conflicting results on how the chaoticity parameter $\kappa$ in the kicked top influences the presence of quantum temporal periodicity \cite{Ruebeck_Pattanayak_2017}. In addition we establish a relationship between our kicked top periodicities and the quantum resonances identified in the kicked rotor. Our results highlight the complex nature of quantum chaos and challenge typical notions of quantum-classsical correspondence.

\section{Background} \label{background}
\subsection{Kicked Top Model}
The quantum kicked top (QKT) is a finite-dimensional dynamical model used to study quantum chaos, known for its compact phase space and parameterizable chaoticity structure \cite{haake1987}.  The time-dependent, periodically-driven system is governed by the Hamiltonian
\begin{equation}
 H = \hbar\frac{p J_{y}}{\tau} +  \hbar \frac{\kappa J_{z}^2}{2j} \sum_{n=-\infty}^{\infty} \delta (t-n\tau),
\end{equation}
where $\{ J_{x}, J_{y}, J_{z}\}$ are the generators of angular momentum: $[J_i, J_j] = i\epsilon_{ijk} J_k$.  It describes a spin of size $j$ precessing about the $y$-axis together with impulsive state-dependent twists about the $z$-axis with magnitude characterized by the chaoticity parameter $\kappa$.  The period between kicks is $\tau$, and $p$ is the amount of $y$-precession within one period.  The associated Floquet time evolution operator for one period is
\begin{equation}\label{eq:floquet-unitary}
    U= \exp\Big(-i\frac{\kappa }{2j}J_{z}^2\Big) \exp\Big(-i\frac{p }{\tau}J_{y}\Big)
\end{equation}

The classical kicked top can be obtained by computing the Heisenberg equations for the re-scaled angular momentum generators, $J_i/j$, followed by the limit $j \to \infty$ \cite{haake1987}.  In the commonly considered case of $(\tau=1, p = \pi/2)$, the classical map is 
\begin{eqnarray}
 \nonumber X_{n+1}&=&Z_n\cos(\kappa X_n)+Y_n\sin(\kappa X_n), \\ 
 \nonumber Y_{n+1}&=&Y_n\cos(\kappa X_n)-Z_n\sin(\kappa X_n), \\
 Z_{n+1}&=&-X_n.
\end{eqnarray}

\begin{figure}[!h]
  \centering
  \subfloat[]{\includegraphics[width=0.5\textwidth]{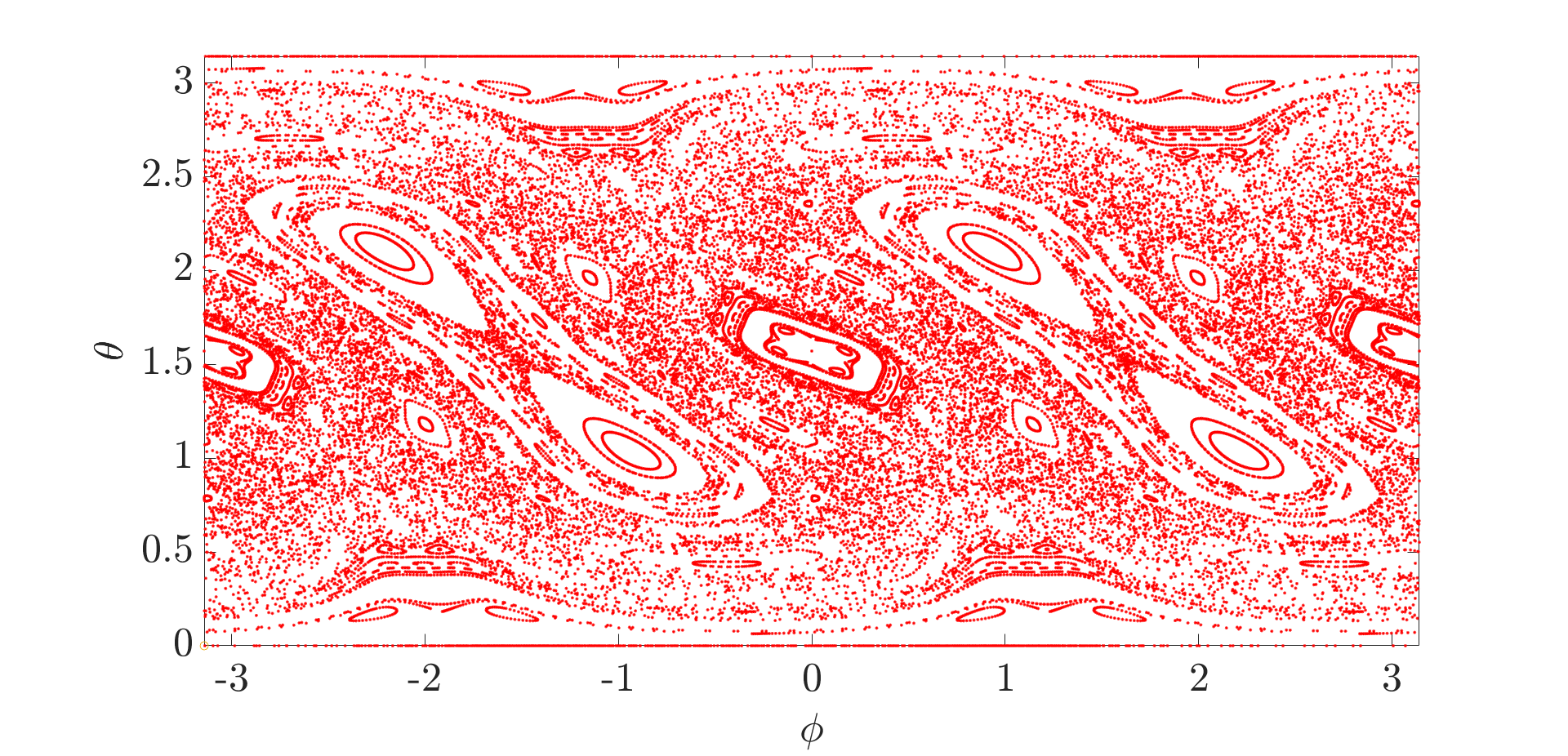}\label{classical_k2.5}}
  \hfill
  \subfloat[]{\includegraphics[width=0.5\textwidth]{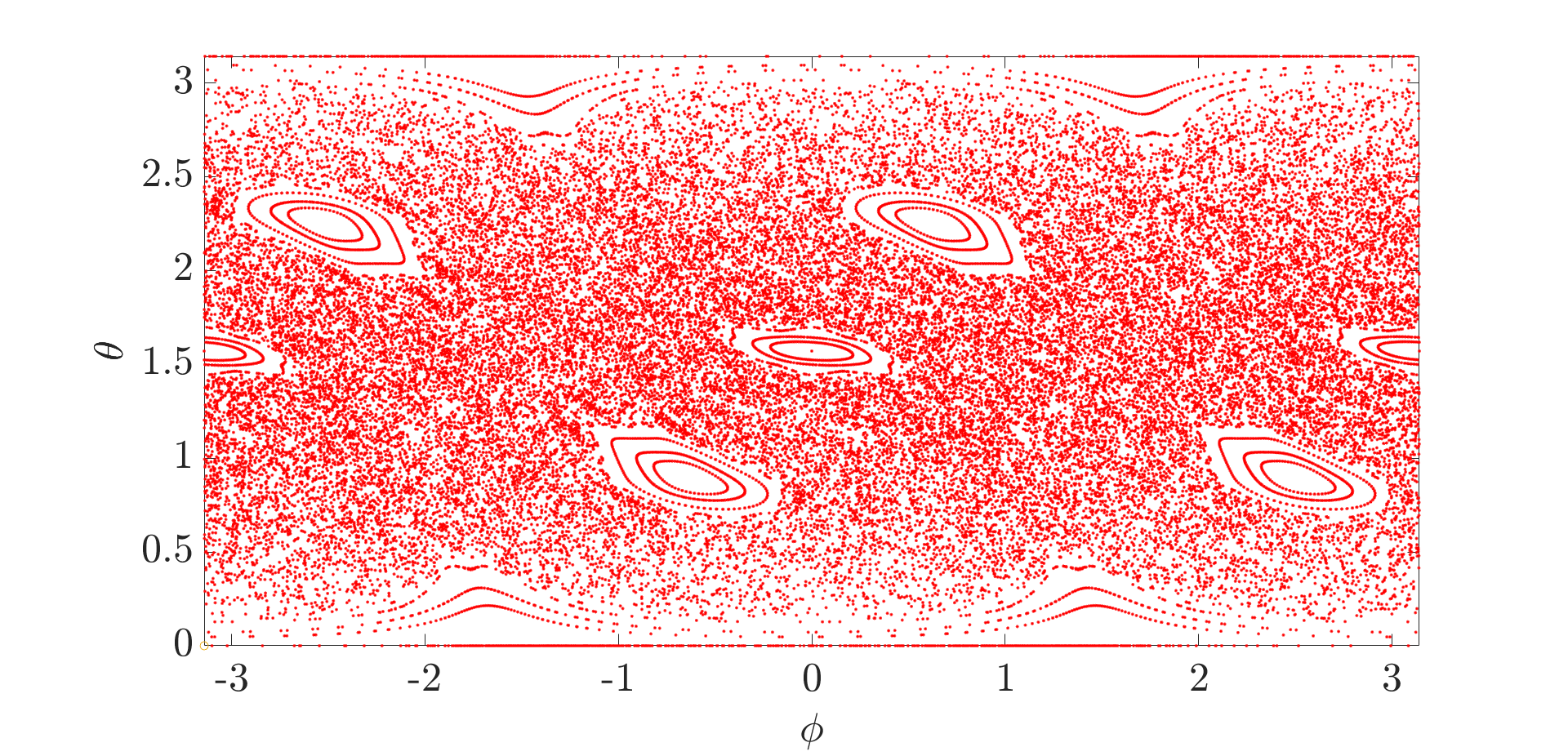}\label{classical_k3}}
  \caption{Stroboscopic map showing the classical time evolution over 150 kicks for \textbf{a}. $\kappa = 2.5$ and \textbf{b}. $\kappa = 3.0$ for several hundred initial points.} 
\end{figure}

As the chaoticity parameter $\kappa$ is varied the classical dynamics ranges from completely regular motion ($\kappa$ $\leq $ 2.1) to a mixture of regular and chaotic motion (2.1 $\leq \kappa \leq 4.4$) to fully chaotic motion ($\kappa > 4.4$) \cite{Kumari_Ghose_vicinity_2018}.  The classical stroboscopic map in polar coordinates for a set of initial conditions with $\kappa= 2.5$ and $\kappa=3.0$ is given in  Fig.\ref{classical_k2.5} and Fig.\ref{classical_k3} respectively.

\subsection{Husimi function}
To study the quantum-classical correspondence in the quantum kicked top, the Husimi function is often used as an aid to compare quantum vs.\ classical dynamics \cite{Kumari_Ghose_vicinity_2018, gs_agarwal_Husumi_1981}.  It is a non-negative quasiprobability distribution defined as
\begin{equation}\label{eq:Husimi_def}
    Q_\rho(\theta,\phi) := \bra{\theta,\phi}\rho\ket{\theta,\phi},
\end{equation}
subject to the normalization condition
\begin{equation}
    \frac{2j+1}{4\pi} \int_{S^2} Q_\rho(\theta,\phi) \sin\theta d\theta d\phi = 1,
\end{equation}
where $\ket{\theta,\phi}$ are the standard spin coherent states associated with SU(2) dynamical symmetry \cite{kulish_1983_quantum}.

\section{Periodicity in twist strength} 
\label{kappa periodicity}

As pointed out in \cite{Bhosale_Santhanam_periodicity_2018}, there is a recurrent relationship between unitaries separated by an amount $\kappa=2\pi j$:
\begin{align*}
U_{\kappa + 2\pi j} &= e^{-i\frac{(\kappa + 2\pi j)}{2j}J_z^2}e^{-ipJ_y} = e^{-i\pi J_z^2} U_\kappa.
\end{align*}
The unitary $e^{-i\pi J_z^2}$ characterizes the difference between the actions of $U_k$ and $U_{k + 2\pi j}$ on Hilbert space.  We will show that this operator acts as a symmetric local unitary in the qubit picture and so does not modify any correlations between the qubits.

Denoting $Z_k := \sigma_z^{(k)}$, consider the operator $e^{-i\pi J_z^2}$ in the qubit picture:
\begin{align}
    e^{-i\pi J_z^2} &= \exp\Big[ -i \frac{\pi}{4} (Z_1 + \cdots + Z_N)^2 \Big] \nonumber \\
    &= \exp\Big[ -i \frac{\pi}{4} \sum_{\vec{k}} \binom{2}{\vec{k}} Z_1^{k_1}\cdots Z_n^{k_n} \Big] \\
    &= \prod_{\vec{k}} \exp\Big[ -i\frac{\pi}{4} \binom{2}{\vec{k}} Z_1^{k_1}\cdots Z_n^{k_n} \Big], \nonumber
\end{align}
where $\vec{k} = (k_1,...,k_n)$ is a multi-index of positive integers that sums to 2, and $\binom{2}{\vec{k}}$ is a multinomial coefficient.  Separate the multi-indices into those with a single $k_i=2$ and those that don't; the former will happen $n$ times, and the associated Pauli operator squares to the identity:
\begin{align}
    \exp\Big[ -i \frac{\pi}{4} I \Big]^n \prod_{\vec{k}\neq 2} \exp\Big[ -i\frac{\pi}{4} \binom{2}{\vec{k}} Z_1^{k_1}\cdots Z_n^{k_n} \Big].
\end{align}
The remaining indices each have exactly two different slots equal to 1 and so the multinomial coefficient is always 2.  The exponentials consequently reduce to
\begin{align}\label{eq:2jpi-intermediate}
    & \,\,\quad e^{-i\frac{n\pi}{4}} \prod_{\vec{k}\neq 2} \left( I^{\otimes n} \cos \frac{\pi}{2} - i Z_1^{k_1}\cdots Z_n^{k_n} \sin \frac{\pi}{2}  \right) \nonumber \\
    &= e^{-i\frac{n\pi}{4}} \prod_{\vec{k}\neq 2} \left( - i Z_1^{k_1}\cdots Z_n^{k_n}\right). 
\end{align}
It is already clear from Eq.\ \eqref{eq:2jpi-intermediate} that $e^{-i \pi J_z^2}$ is a local unitary and so does not affect any correlations between the qubits.  

Hence, the entanglement generated between the qubits is periodic in the chaoticity parameter with period $\Delta \kappa = 2\pi j$ \cite{Bhosale_Santhanam_periodicity_2018}.  That is to say,
\begin{equation}\label{eq:floquet-2pij-lo}
   U_{\kappa + 2\pi j} \overset{\text{LO}}
   {\Longleftrightarrow} U_{\kappa},
\end{equation}
where LO refers to (symmetric) local operations over the global Hilbert space of the qubits.

Eq.\ \eqref{eq:2jpi-intermediate} can be written in more compact form as
\begin{equation}\label{eq:pi_twist_symmetric}
    e^{-i \pi J_z^2 } = (-1)^{j^2} Z_1^{n-1}\cdots Z_n^{n-1},
\end{equation}
which is clearly symmetric.  This breaks into three cases of spin
\begin{equation}\label{eq:pi-twist-qubit-picture}
    e^{-i \pi J_z^2 } = \begin{cases}
    Z^{\otimes n} &  \text{even integer} \\
    -Z^{\otimes n} &  \text{odd integer} \\
    e^{-i \frac{\pi}{4}} I^{\otimes n} & \text{half-integer}
\end{cases}.
\end{equation}

\section{Temporal periodicity}
Here we derive the temporal periodicity of the kicked top evolution for three special values of twist strength: $\{2\pi j, \pi j, \frac{\pi j}{2}\}$, each of which are split into cases of integer and half-integer spins.

\subsection{Twist strength \texorpdfstring{$\kappa = 2\pi j$}{TEXT}}\label{sec:2pij_integer}
The Floquet operator in the case of $\kappa = 2\pi j$ is 
\begin{equation}\label{eq:}
    U_{2\pi j} = e^{-i \pi J_z^2} e^{-i p J_y},
\end{equation}
where $e^{-i\pi J_z^2}$ is a symmetric local unitary in the qubit picture \eqref{eq:pi_twist_symmetric}.  Like many of the results here, the consequences on temporal periodicity strongly depends on whether the spin is integer or half-integer.

\subsubsection*{Integer spin}
In the case of integer spin the evolution squares to the identity regardless of the $y$-rotation angle: 
\begin{equation}\label{eq:2jpi_integer_period2}
    U^2_{2\pi j} = I^{\otimes n}.    
\end{equation}
This can be seen using either integer form of Eq.\ \eqref{eq:pi-twist-qubit-picture} and writing the $y$-rotation in the qubit picture as
\begin{equation}\label{eq:y-rotation-qubit-picture}
    e^{-i p J_y} = ( I \cos \frac{p}{2} - i Y \sin \frac{p}{2} )^{\otimes n}.
\end{equation}
And because in this case $U^2_{2\pi j}$ is simply a composition of symmetric local unitaries, it suffices to consider a single tensor factor:
\begin{equation}
\begin{split}
    &\quad [ Z ( I \cos \frac{p}{2} - i Y \sin \frac{p}{2} )]^2 \\
    &= [ Z \cos \frac{p}{2} - X \sin \frac{p}{2} ]^2 \\ 
    &= I(\cos^2\frac{p}{2} + \sin^2\frac{p}{2}) - \frac{1}{2}(ZX + XZ)\sin p \\
    &=I
\end{split}    
\end{equation}
with the last line coming from the anti-commutation relations of Pauli matrices.

\begin{figure}[h!]
  \centering
  \includegraphics[width=0.45\textwidth]{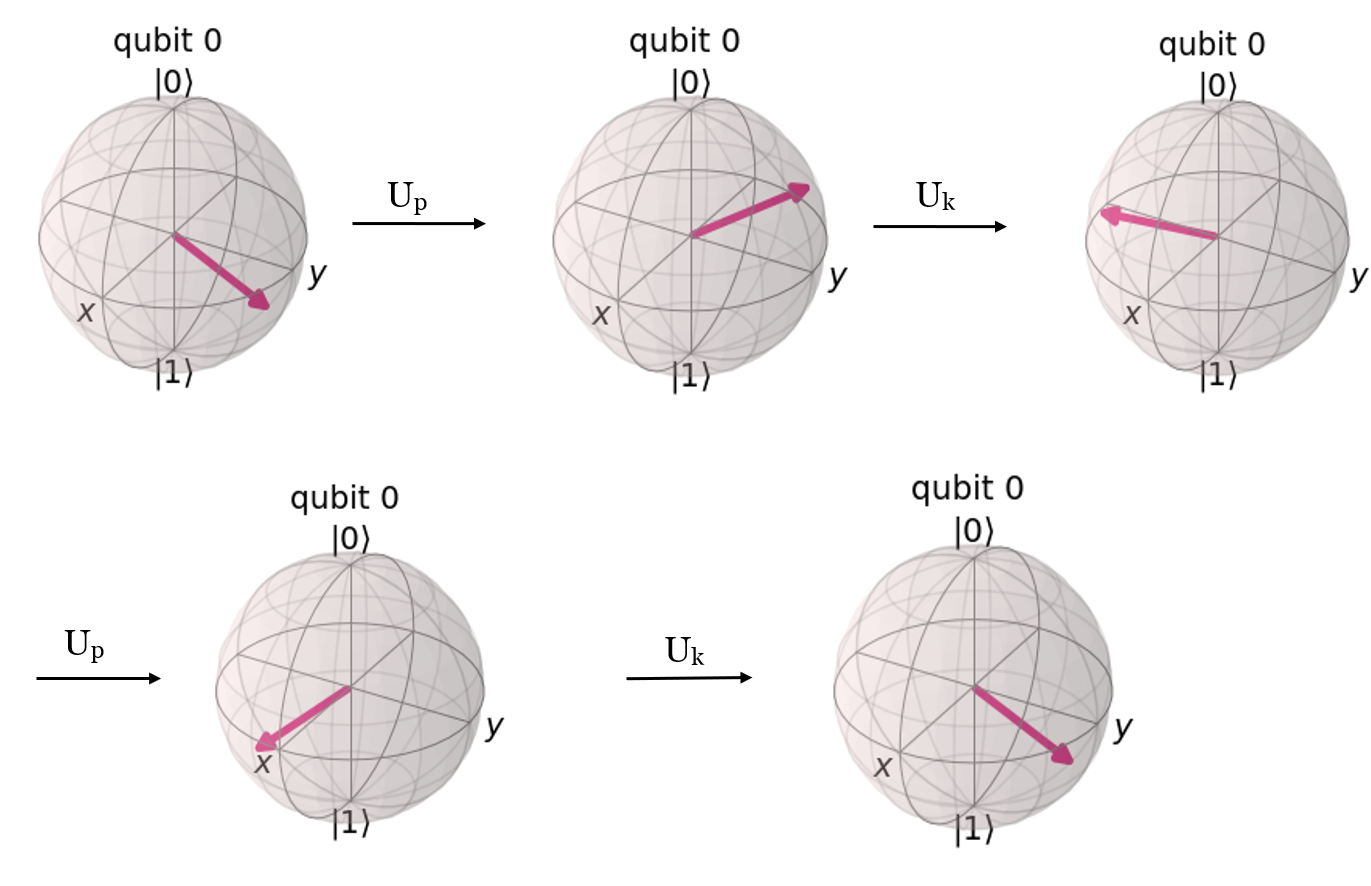}
  \caption{Evolution of the Bloch vector associated to any reduced qubit state for $\kappa = 2\pi j$.  After two kicks the state returns to its initial point, showing a two-step temporal periodicity.  The initial point is $(\theta,\phi) = $ (2.25,2.0).} 
  \label{state evloution 2jpi}
\end{figure}

See Fig.\ \ref{state evloution 2jpi} for a visual interpretation of Eq.\ \eqref{eq:2jpi_integer_period2} by tracking the collectively shared Bloch vector.  After the first $y$-rotation the twist effectively acts as a $\pi$-rotation about the $z$ axis.  Consequently, the second $y$-rotation then undoes the first and the second twist rotates back to the starting point.  Note that this demonstration depends on each step being a symmetric local unitary, meaning that a spin coherent state will remain so throughout and no correlations are ever generated.

\subsubsection*{Half-integer spin}

In the case of half-integer spin the twist becomes a scaled identity operator \eqref{eq:pi-twist-qubit-picture}, leading to a $p$-dependent temporal periodicity (up to an irrelevant global phase) of $N$ kicks under the condition
\begin{equation}
    U_{2\pi j}^N = I \quad \text{if} \quad N = \frac{2\pi}{p} \in \mathbb{N}
\end{equation}
Hence only when $p$ is a rational fraction of $\pi$ does there exist a temporal periodicity at this twist strength.  It is interesting that in the half-integer case the rotation angle $p$ is critical for determining the existence of a temporal periodicity while in the integer case $p$ has no effect.

\subsection{Twist strength \texorpdfstring{$\kappa = \pi j$}{TEXT}}
The case of twist strength $\kappa = \pi j$ yields a more interesting temporal periodicity that also depends on the spin being integer or half-integer.  Details of calculations can be found in Supplementary \ref{supplementry-2}.

\subsubsection*{Integer spin}
Following a similar argument from the previous section, the general expression for the twist unitary $e^{-i\frac{\kappa}{2j}J_z^2}$ is
\begin{equation}\label{eq:general-twist-qubit}
    e^{-i\frac{\kappa}{8j}n} \prod_{\vec{k}\neq 2} \left( I^{\otimes n} \cos \frac{\kappa}{4j} - i Z_1^{k_1}\cdots Z_n^{k_n} \sin \frac{\kappa}{4j}  \right),
\end{equation}
which reduces to 
\begin{equation}
   e^{-i\frac{\pi}{2}J_z^2} = e^{-i\frac{\pi}{8}n} \prod_{\vec{k}\neq 2} \frac{1}{\sqrt{2}} \left( I^{\otimes n} - i Z_1^{k_1}\cdots Z_n^{k_n}  \right)
\end{equation}
when $\kappa = \pi j$.  This appears to be a difficult expression to evaluate but simplifies to
\begin{equation}\label{eq:twist-jpi-integer}
    e^{-i\frac{\pi}{2}J_z^2} = e^{-i\frac{\pi}{4}}\frac{I^{\otimes n} + i (iZ)^{\otimes n}}{\sqrt{2}},
\end{equation}
as can be verified by comparing the two actions on the computational basis in $(\mathbb{C}^2)^{\otimes 2j}$.  This can also be found using the Gaussian sum decomposition result from \cite{Zou_Wang_pseudo_2022}.  With this in mind, and writing the $y$-rotation as in Eq.\ \eqref{eq:y-rotation-qubit-picture}, the Floquet operator can be shown to exhibit the finite-time periodicity
\begin{equation}\label{eq:j-pi-integer-spin-8th-power}
    U_{\pi j}^8 = I \qquad \forall \, \, \text{integer } \, j.
\end{equation}

This can be done by establishing
\begin{equation}\label{eq:pi_j_int_U4}
\begin{split}
    U_{\pi j}^4 &= \left[ \frac{e^{-i\frac{\pi}{4}}}{\sqrt{2}}\bigg(I^{\otimes n} + i (iZ)^{\otimes n}\bigg) \bigg(\frac{I - i \sigma_y }{\sqrt{2}}  \bigg)^{\otimes n} \right]^4 \\
    &= -(iY)^{\otimes n}
\end{split}
\end{equation}
through repeated use of the Pauli group commutation relations.  As $n$ is an even integer and $Y^2 = I$, this is enough to give Eq.\ \eqref{eq:j-pi-integer-spin-8th-power}.  The same calculation may be repeated for $\kappa = \pi j + 2\pi j = 3\pi j$ which also shows the period 8 periodicity.  While expected from the $2\pi j$ periodicity in correlations \cite{Bhosale_Santhanam_periodicity_2018}, this additional calculation is necessary to conclude the stronger notion of temporal periodicity of the state itself, possibly up to global phase.  For example, any SU(2) rotation with an angle incommensurate to $\pi$ will produce a sequence of spin coherent states  -- and therefore a period-1 recurrence in the quantum correlations -- that never returns to the original state exactly.

\begin{figure}[h]
  \centering
  \includegraphics[width=\columnwidth]{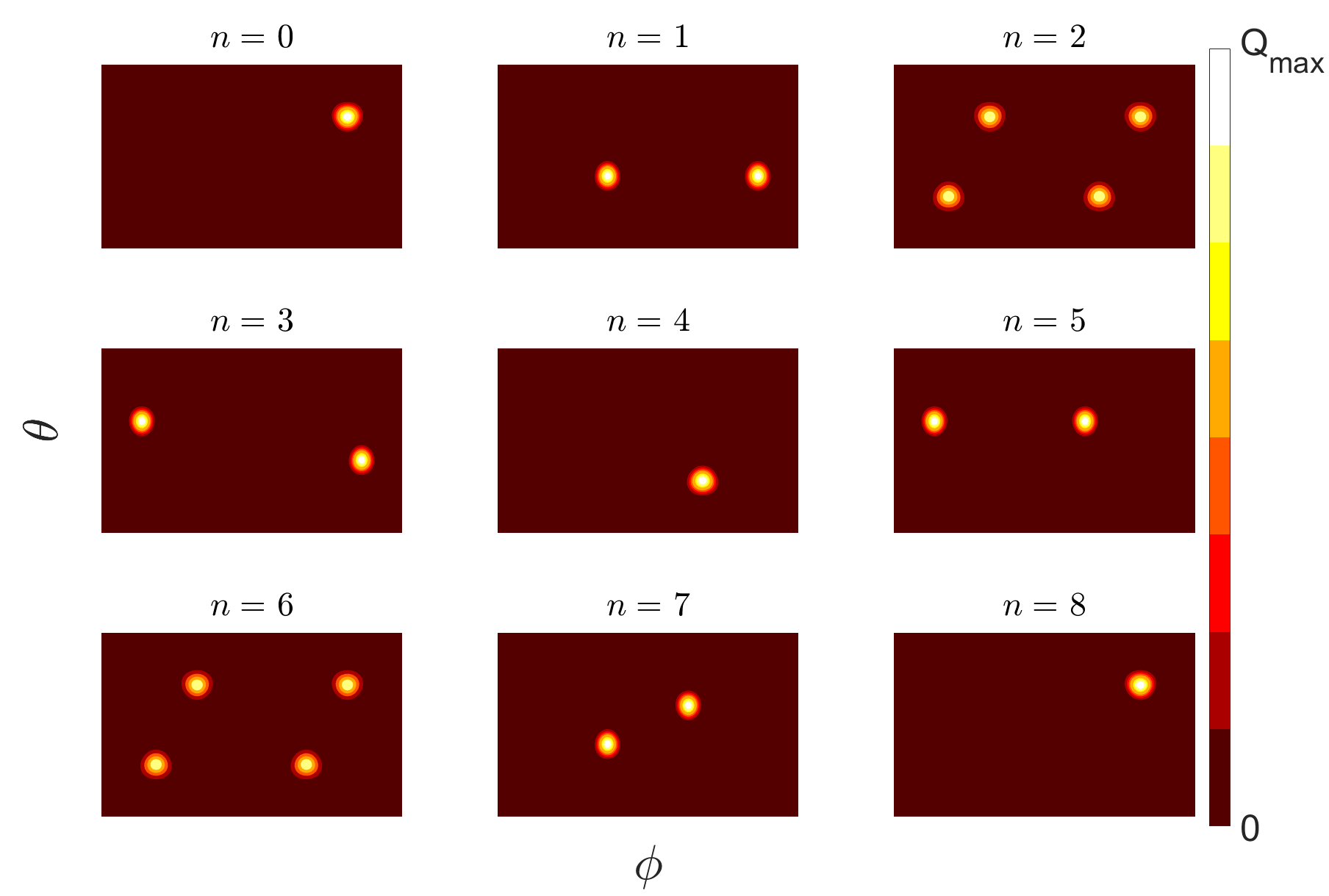}
  \caption{Stroboscopic Husimi evolution at $\kappa=\pi j$ of a spin coherent state starting at $(\theta,\phi) = (2.25,2.0)$ over 8 kicks.  The state splits and becomes entangled then recombines back to the original unentangled position. $Q_{max}$ corresponds to the maximum height of the Husimi distribution in each plot.  Here $j=50$.} 
  \label{fig:integer_j_Husimi_evolution_jpi}
\end{figure}

In contrast to the previous twist strength of $\kappa=2\pi j$, here entanglement is generated (and destroyed) throughout the period-8 orbit.  This can be seen from Eq.\ \eqref{eq:twist-jpi-integer} which is clearly not a symmetric local unitary.  Fig.\ \ref{fig:integer_j_Husimi_evolution_jpi} shows the orbit in the Husimi representation \eqref{eq:Husimi_def} of a $j=50$ spin coherent state initially centred at $(\theta=2.25,\phi=2.0)$.  After the initial rotation about the $y$-axis, we see the action of \eqref{eq:twist-jpi-integer} ``splitting'' the state into a cat-like superposition.  A second round of rotation-twist iteratively produces a balanced superposition of four spin coherent states distributed over phase space.  Another two kicks recombines this state into the initial spin coherent state but reflected about the y-axis, matching \eqref{eq:pi_j_int_U4}.  Finally another four kicks repeats this process, resulting in a recurrence of the initial state.  
This regular, periodic dynamical behaviour appears to have no analogue in the classical kicked top (not least of which at $\kappa=\pi j$) and so represents a departure from the classical-quantum correspondence.

It should also be noted that while the above is the generic temporal periodicity, certain states related to the Hamiltonian symmetries will experience a shorter orbit.  In particular if we take the initial state as $\ket{+}_y$, i.e.\ $(\theta,\phi) = (\pi/2,\pi/2)$, then the rotation part of the unitary will be ineffective.  The twist \eqref{eq:twist-jpi-integer} will create the superposition of $\ket{+}_y$ and $\ket{-}_y$; it can be shown that the evolution reduces to a period-4 orbit for even integer spins and a period-2 orbit for odd integer spins.  See also \cite{Madhok_Dogra_Lakshminarayan_2018} for a related analysis. 

\subsubsection*{Half-integer}
In the case of half-integer spin and $\kappa = j\pi$ the general twist operator, Eq.\ \eqref{eq:general-twist-qubit}, is equivalent to the following unitary in the qubit picture:
\begin{align}
    e^{-i\frac{\pi}{2}J_z^2} &= e^{-i\frac{\pi}{8}} \frac{1}{\sqrt{2}} \left[  R^\dagger_z(\frac{\pi}{2}) + R_z(\frac{\pi}{2})  \right]  \\ 
    &= \frac{e^{-i\frac{\pi}{8}}}{\sqrt{2}}\Bigg[\left(\frac{I + i Z}{\sqrt{2}}\right)^{\otimes n} + \left(\frac{I - i Z)}{\sqrt{2}}\right)^{\otimes n} \Bigg]. \label{eq:twist-jpi-half-integer}
\end{align}

Similar to the integer case, repeated and iterated use of the Pauli group commutation relations show that Eq.\ \eqref{eq:twist-jpi-half-integer} raised to the 6th power yields a $\pi$-rotation up to phase:

\begin{equation}
    U^{6}_{\pi j} = e^{-i\frac{3\pi}{4}}(iY)^{\otimes n}.
\end{equation}
The full state recurrence comes after 12 kicks:
\begin{equation}
    U_{\pi j}^{12} = e^{-i\frac{\pi}{2}}I \qquad \forall \, \, \text{half-integer } \, j,
\end{equation}
which is a finite temporal periodicity up to global phase.  To our knowledge this recurrence was first discovered for the spin-$\frac{3}{2}$ case in Ref.\ \cite{Dogra_exactly_2019}; here we have shown that it exists in all dimensions.  Also, as expected, a similar calculation shows another period-12 recurrence at $\kappa = \pi j + 2\pi j = 3\pi j$, similar to the integer-spin case.

Again starting with a generic spin coherent state (i.e.\ a symmetric product state in the qubit picture), entanglement is generated and destroyed throughout its 12-state orbit.  Similar to Fig.\ \ref{fig:integer_j_Husimi_evolution_jpi}, the generation occurs during the recursive splitting of the state into successive cat-like superpositions, and the destruction occurs during the subsequent recombination into new, displaced spin coherent states.  States associated with Hamiltonian symmetries again experience a reduced orbit length. For the initial state as $\ket{+}_y$, i.e.\ $(\theta,\phi) = (\pi/2,\pi/2)$, the evolution reduces to a period-3 orbit.

\subsection{Twist strength \texorpdfstring{$\kappa = \frac{\pi j}{2}$}{TEXT}}
The case of $\kappa = \frac{\pi j}{2}$ has the most apparent difference between integer and half-integer spin.

\subsubsection*{Integer spin}
Using the Gaussian sum decomposition \cite{Zou_Wang_pseudo_2022}, for integer spin the twist operator $e^{-i\frac{\pi}{4}J_z^2}$ splits into the superposition of rotations
\begin{equation}
    \frac{1}{2}\Big[ e^{-i\frac{\pi}{4}}I + e^{-i\frac{\pi}{2}J_z} + e^{i\frac{3\pi}{4}}e^{-i\pi J_z} + e^{-i\frac{3\pi}{2}J_z} \Big].
\end{equation}
In the qubit picture this becomes
\begin{equation}
    \begin{split}
        e^{-i\frac{\pi}{4}J_z^2} &= \frac{1}{2}\Bigg[ e^{-i\frac{\pi}{4}} I^{\otimes n} + \left( \frac{I - iZ}{\sqrt{2}} \right)^{\otimes n}  \\
        &\qquad + e^{i\frac{3\pi}{4}}\left( iZ \right)^{\otimes n} + \left( \frac{I + iZ}{\sqrt{2}} \right)^{\otimes n} \Bigg]
    \end{split}
\end{equation}
where we have used the fact that $n$ is even to simplify.  Numerical calculations suggest a temporal periodicity with period 48,
\begin{equation}
    U^{48}_{\frac{\pi j}{2}} = I \qquad \forall\text{ integer }j.
\end{equation}
This has been confirmed up to spin $j=500$ where the Hilbert–Schmidt distance $\| U^{48}_{\frac{\pi j}{2}} - I \|$ remains zero within the working error tolerance of $10^{-10}$.  And similar to the previous $\kappa=\pi j$ case (both integer and half-integer) here the Floquet operator raised to half the periodicity (i.e.\ 24) also acts as an effective $\pi$-rotation about the $y$-axis up to some global phase.  Cat-like splitting and recombination cycles were furthermore observed in the Husimi function tracking of a generic spin coherent state.  Part of the difficulty in showing this analytically comes from determining the twist operator in the qubit picture as in Eqs.\ \eqref{eq:twist-jpi-integer} and \eqref{eq:twist-jpi-half-integer}.  Numerics also confirm a period-48 recurrence at $\kappa = \frac{\pi j}{2} + 2\pi j = \frac{5\pi j}{2}$.

We also note what appears to be two higher-frequency temporal recurrences present in low dimensions at this chaoticity value: for $j=1$ and $j=3$ the evolution repeats after only 16 kicks rather than 48.  This observation is distinct from the continued theme of the special states $\ket{\pm}_y$ experiencing a reduced orbit of 24 for even values of $j$ and 4 for odd values of $j$, which we numerically verified.

\subsubsection*{Half-integer spin}
 In the half-integer case we surprisingly find no temporal periodicity for $\kappa = \frac{\pi j}{2}$.  This was numerically concluded by computing the entanglement entropy of any one of the reduced constituent qubits,
\begin{equation}
    \rho = \frac{1}{2}
    \begin{pmatrix}
    1 - \langle S_z \rangle & \langle S_- \rangle \\
    \langle S_+ \rangle & 1 + \langle S_z \rangle
    \end{pmatrix},
\end{equation}
via the collective spin observables $\{S_z, S_\pm = S_x \pm i S_y \}$ where $S_i = J_i/j$ \cite{Baguette_MM_reductions_2014}.  This approach was used instead of Hilbert-Schmidt distance to avoid optimizing over the angles $\varphi$ that could have \textit{a priori} appeared in a hypothetical periodicity of the form $U^n = e^{i\varphi} I$.  

\begin{figure}
  \centering
  \includegraphics[width=1.05\columnwidth]{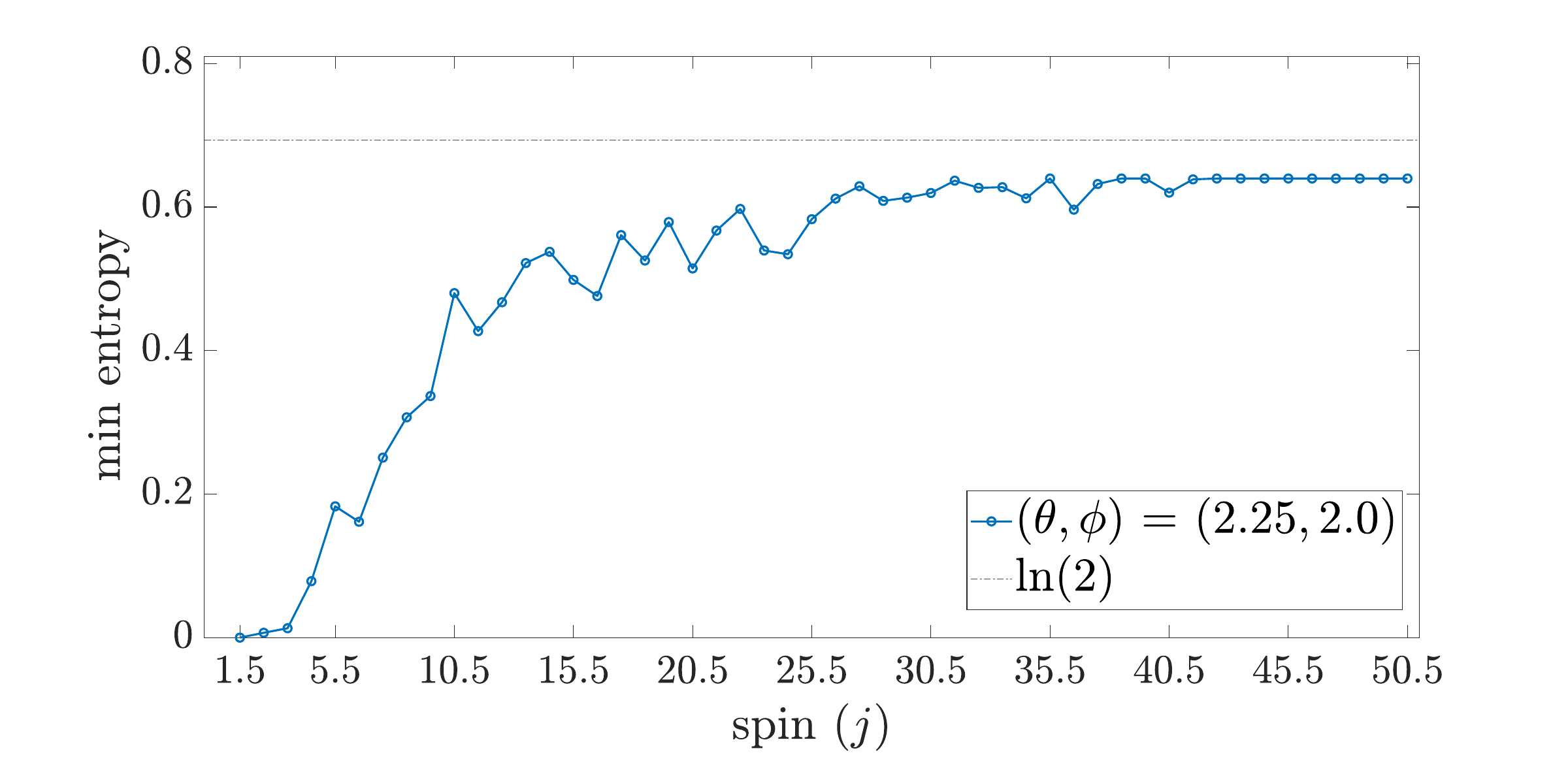}
  \caption{Minimum single-qubit entropy within the first $5000$ kicks at $\kappa=\frac{\pi j}{2}$ as a function of spin.  Initial spin coherent state is centred at $(\theta,\phi) = (2.25,2.0)$.} 
  \label{min_entropy}
\end{figure}

All that is needed to conclude the lack of a global recurrence is the identification of a spin coherent state that never returns to product form.  We thus focus on our running example of $\ket{\theta,\phi}=\ket{2.25, 2.0}$.  We found that up to spin $j=50\frac{1}{2}$, the single qubit dynamical entropy never falls below $10^{-5}$ within the first 5000 kicks.  In fact, the entropy generally increased with dimension.  Fig.\ \ref{min_entropy} plots the \textit{smallest} entanglement entropy obtained by any of qubits throughout the first 5000 kicks.  As can be seen, higher spins experience a highly entangled orbit, remaining close to the upper bound of $S_{\text{max}} = \ln{2}$.

Further evidence supporting the lack of a recurrence can be found in the specific case of spin $j=\frac{3}{2}$, the smallest possible kicked top applicable to this scenario. Recently, many aspects of this low-dimensional system were solved exactly, including the single-qubit linear entropy
\begin{equation}
    S^{(\text{lin})}_\rho = 1 - \Tr[\rho^2]
\end{equation}
of various initial spin states as a function of twist strength and kick number $N$ \cite{Dogra_exactly_2019}.  In particular, the single-qubit linear entropy of the state $U^N_{\kappa}\ket{+}_y$ was found to be
\begin{equation}\label{eq:3_qubit_+y_lin_end}
    S^{(\text{lin})}(N,\kappa) = 4\chi^2 U^2_{N-1}(\chi)[ 1 - 2\chi^2 U^2_{N-1}(\chi) ],
\end{equation}
where
\begin{equation}
    U_{N-1}(\chi) = \frac{\sin(N\gamma)}{\sin(\gamma)}
\end{equation}
are the Chebyshev polynomials with arguments related to the twist strength via
\begin{equation}
    \chi = \cos(\gamma) = \frac{1}{2}\sin(\frac{\kappa}{3}).
\end{equation}
Eq.\ \eqref{eq:3_qubit_+y_lin_end} may be efficiently computed using symbolic programming and we found that the linear entropy does not exactly vanish within the first million kicks.

Given that a recurrence is almost certainly not present in the $j=3/2$ system at this twist strength, it seems highly unlikely that a family of recurrences exist, one for each half-integer $j>3/2$. This argument has an added strength by focusing on the special state $\ket{+}_y$, which, due to the Hamiltonian symmetry of the system, has a pattern of experiencing a reduced recurrence time when a global periodicity exists.

The lack of periodicity at this $\kappa$ value also shows that in general, not all twist strengths commensurate to $\pi$ yield an exact recurrence.

\subsection{Summary and other resonances}
Table \ref{summary_table} summarizes our results.  With these recurrences established, a natural question to ask is if there are others.  To this end, we have performed a numerical search for such recurrences characterized by $\kappa = \pi j \frac{r}{s}$ for all coprime $1 \leq r,s \leq 10$, for all integer and half-integer spin, upto $15.5$ and have found none. This was done by computing the von-Neuman entropy for the initial state given by $\ket{\theta,\phi}=\ket{2.25, 2.0}$, upto 500 kicks and found that minimum value of von-Neuman entropy for the different sets of r and s upto  $j=15.5$ never falls below $10^{-7}$. Numerical simulations suggests that there are no other sets of $r$ and $s$ that shows the temporal periodicity, than what we 
 have found. This therefore places constraints on any additional values of $\kappa$ that yield a state-independent finite periodicity.

\newcommand\xrowht[2][0]{\addstackgap[.5\dimexpr#2\relax]{\vphantom{#1}}}
\begin{table}[h]
\centering

\begin{tabular}{|c|c|c|}
\hline\xrowht[()]{7pt}
\textbf{ chaos parameter   } & \multicolumn{2}{c|}{\textbf{period}}\\
\hline\xrowht[()]{7pt}
 $\mathbf{\kappa}$ & integer spin & half-integer spin \\
\hline\xrowht[()]{7pt}
 0 & 4 & 4 \\
\hline\xrowht[()]{7pt}
 $\frac{\pi j}{2}$ & $48^*$ & $\times^*$ \\
\hline\xrowht[()]{7pt}
 $\pi j$ & 8 & 12\\
\hline\xrowht[()]{7pt}
 $\frac{3\pi j}{2}$  & $48^*$ & $\times^*$ \\
\hline\xrowht[()]{7pt}
$ 2\pi j$ & 2 & 4 \\
\hline\xrowht[()]{7pt}
$ \frac{5\pi j}{2}$  & $48^*$ & $\times^*$ \\
\hline\xrowht[()]{7pt}
$ 3\pi j$ & 8 & 12 \\
\hline\xrowht[()]{7pt}
$ \frac{7\pi j}{2}$  & $48^*$ & $\times^*$ \\
\hline\xrowht[()]{7pt}
$ 4\pi j$ & 4 & 4 \\
\hline
\end{tabular}
\caption{Recurrence periods for different $\kappa$ values. Here $\times$ signifies the non-existence of periodicity and (*) represents results from numerical simulation.  The numbers are specific to $p=\frac{\pi}{2}$ with the exception of the integer-spin period-2 orbit for $\kappa=2\pi j$, which is independent of $p$.}
\label{summary_table}
\end{table} 

\subsection{Relation to kicked rotor}
It is interesting to compare our results to the quantum resonance behaviour found in the quantum kicked rotor \cite{izrailev_1980_quantum,fishman_rotor_anderson_1982,kanem_2007_observation}.  This purely quantum dynamics occurs when one of the  Hamiltonian parameters takes the form $4\pi\frac{r}{s}$ and is characterized by quadratic growth of the wavefunction in momentum space.  In contrast, the classical kicked rotor at the same parameter value only has linear scaling.  An interesting exception to this quadratic growth behaviour is the case of $r/s = 1/2$, which yields a period-2 state-independent orbit.  This special case is known as \textit{quantum anti-resonance} due to the complete lack of momentum growth \cite{izrailev_1980_quantum}.

Ref.\ \cite{Zou_Wang_pseudo_2022} proposed a kicked top version of the quantum resonance condition as
\begin{equation}\label{eq:qkt_resonance}
    \kappa = 4\pi \frac{r}{s} j
\end{equation}
for coprime integers $r$ and $s$, where it may be assumed $r/s < 1$ without loss of generality due to the global symmetry $U_{\kappa} = U_{\kappa + 4\pi j}$.  This proposal is motivated by the well-known contraction from the quantum kicked top to the quantum kicked rotor \cite{Haake_limit_1988}, effected via the simultaneous scaling
\begin{equation}\label{eq:contraction_scaling}
    \kappa \sim j \qquad p \sim \frac{1}{j} \quad \text{ as } \quad j \rightarrow \infty.
\end{equation}
Here the kicked top parameter $\kappa$ becomes the relevant parameter in the kicked rotor that controls the existence of resonances.  (Also note that despite $j\rightarrow\infty$ the above is not to be considered a classical limit as the quantum kicked rotor is a fully quantum object -- hence \textit{contraction}.)

The periodicities examined here do not satisfy the $p \sim 1/j$ scaling \eqref{eq:contraction_scaling} and therefore are not to be seen as ``pre-contracted'' phenomena, at least not in a strict sense.  It is thus interesting that despite only having a partial relationship to the resonance behaviour found in the kicked rotor we still observe non-standard dynamics in the kicked top at these special chaoticity values.

The lone case of $\kappa = 2\pi j$ (i.e.\ $r/s=1/2$) for integer spin discussed in sec.\ \ref{sec:2pij_integer} actually can be seen as being ``pre-contracted''.  This is because the period-2 orbit does not depend on the rotation angle $p$, and so without loss of generality we may set it to scale as $p \sim 1/j$.  Thus the peculiar behaviour of quantum anti-resonance found in the rotor may be seen as originating in the quantum kicked top.  Previous works focusing on quantum correlations \cite{Bhosale_Santhanam_periodicity_2018} or the pseudo-classical framework \cite{Zou_Wang_pseudo_2022} do not fully capture this specialized anti-resonance effect: both approaches depend on the rotation angle $p$ and both predict a higher than necessary orbit period \footnote{Ref.\ \cite{Bhosale_Santhanam_periodicity_2018} predicts a period-2 orbit in the quantum correlations at this chaoticity value.  This is not wrong, but as shown in sec.\ \ref{sec:2pij_integer} no entanglement is generated in the intermediate state and therefore the quantum correlations actually experience a period-1 orbit while the state experiences a period-2 orbit.}.

\section{Conclusion}

Previous studies comparing classical and quantum dynamics in the kicked top largely validate the correspondence principle in the semiclassical regime \cite{Ghose_Sanders_ent_dynamics_2004,wang_ghose_entg_2004}. Other works have gone into characterizing if and when the correspondence principle may be applied in the deep quantum regime \cite{anand2021simulating, Neill_2016, Krithika_NMR_QKT_2019, Ruebeck_Pattanayak_2017, Bhosale_Santhanam_periodicity_2018}.  Here we have demonstrated a general violation of the correspondence principle by finding various sets of state-independent, finite-time periodicities that have no classical analog, and which exist for all spins (i.e.\ both the deep quantum and semiclassical regimes).  Some of these recurrences had been identified earlier for specific spins \cite{Dogra_exactly_2019, Ruebeck_Pattanayak_2017} or in a semiclassical context \cite{Zou_Wang_pseudo_2022}, but here we have
generalized these results. We have analytically shown the existences of sets of recurrences and numerically introduced others.  A preliminary search for additional ``simple'' periodicities indicate that if they exist the recurrence time must be relatively large.

Our analysis resolves a confusion over the general relationship between the rationality of the chaoticity parameter $\kappa$ and the existence of a recurrence in the quantum kicked top. Ref.\ \cite{Ruebeck_Pattanayak_2017} argued that whenever this value is a rational multiple of $\pi$ the evolution will be periodic in the sense that any initial state will only explore a finite subset of Hilbert space.  Ref.\ \cite{Bhosale_Santhanam_periodicity_2018} on the other hand maintained that this is only true for spin-1 systems; i.e.\ higher dimensional kicked tops do not experience such finite-orbit periodicity regardless of the chaoticity value.  Here we have  demonstrated the answer lies somewhere in-between. In particular, while recurrences do exist in all dimensions, and these recurrences do come from a rational $\kappa$ value, not all rational $\kappa$ values may yield a recurrence.

We further established a relationship to the quantum resonance phenomenon of the quantum kicked rotor \cite{izrailev_1980_quantum}, and showed that the peculiar anti-resonance effect (i.e.\ $U^2 = I$) in the kicked rotor may have its origins in the quantum kicked top. Given the link to the kicked rotor and the simple, general criterion for periodicity, it would seem reasonable to expect such non-classical resonances to occur in other periodic or kicked systems as well. Future work in this direction would help shed light on the complicated route to correspondence in chaotic systems.

\section*{Acknowledgements}
This work was supported in part by the Natural Sciences and Engineering Research Council of Canada (NSERC).  We acknowledge fruitful discussions with Alan Jamison and Shlok Nahar.  Wilfrid Laurier University and the University of Waterloo are located in the traditional territory of the Neutral, Anishnawbe and Haudenosaunee peoples. We thank them for allowing us to conduct this research on their land. 


\bibliographystyle{naturemag}
\bibliography{ref}

\widetext 
\begin{center}
\textbf{\large Supplementary Material for: Quantum recurrences in the kicked top}
\end{center}
\setcounter{equation}{0}
\setcounter{figure}{0}
\setcounter{table}{0}
\setcounter{page}{1}
\setcounter{section}{0}
\makeatletter
\renewcommand{\theequation}{S\arabic{equation}}
\renewcommand{\thefigure}{S\arabic{figure}}
\renewcommand{\bibnumfmt}[1]{[S#1]}
\renewcommand{\citenumfont}[1]{S#1}

\section{Twist strength \texorpdfstring{$\kappa = j\pi$}{TEXT}}\label{supplementry-2}
 
Here we will shows the proof for Eq.\ \eqref{eq:twist-jpi-integer} and Eq.\ \eqref{eq:twist-jpi-half-integer} by its action on n-qubit state with hamming weight \textit{k} as
\begin{equation}\label{eq:twist-jpi-integer on hamming state}
     e^{-i\frac{\pi}{2}J_z^2}\ket{D^{(k)}_n} = e^{-i\frac{\pi}{2}\frac{{(n-2k)}^2}{4}}\ket{D^{(k)}_n}.    
\end{equation}
where $\ket{D^{(k)}_n}$ is given by
\begin{equation}\label{eq:hamming state}
     \ket{D^{(k)}_n} = \ket{  \underbrace{00\cdots0}_{n-k} \otimes \underbrace{1 \cdots11}_{k} }.    
\end{equation}
We can write the above since any n-qubit state with same hamming weight is the eigen state of $J_z$ with same eigenvalue.

\subsection{Integer spin}

For $\kappa = j\pi$. and integer value of $j$, acting the left hand side of the Eq.\ \eqref{eq:twist-jpi-integer} on n-qubit state gives us,

\begin{equation}\label{eq:equivalent twist-jpi-integer on hamming state}
\begin{split}
    &\quad e^{-i\frac{\pi}{4}}\Bigg[\frac{I^{\otimes n} + i (i\sigma_{z})^{\otimes n}}{\sqrt{2}}\Bigg]\ket{D^{(k)}_n}  \\
    &= e^{-i\frac{\pi}{4}}\Bigg[\frac{I^{\otimes n} + (i)^{n+1} (\sigma_{z})^{\otimes n}}{\sqrt{2}}\Bigg]\ket{D^{(k)}_n}  \\
    &= e^{-i\frac{\pi}{4}}\Bigg[\frac{1 + (i)^{n+1+2k}}{\sqrt{2}}\Bigg]\ket{D^{(k)}_n}
    \end{split}
\end{equation}

If Eq.\ \eqref{eq:twist-jpi-integer} holds, then Eq.\ \eqref{eq:twist-jpi-integer on hamming state} and Eq.\ \eqref{eq:equivalent twist-jpi-integer on hamming state} are equivalent, which implies,

\begin{equation}\label{eq: equivalent expression of twist operator for even qubits}
     e^{-i\frac{\pi}{2}\frac{{(n-2k)}^2}{4}} = e^{-i\frac{\pi}{4}}\Bigg[\frac{1 + (i)^{n+1+2k}}{\sqrt{2}}\Bigg]    
\end{equation}
Since \textit{n} is even, writing $n=2q$, where $q \in {\{1,2,.....\frac{n}{2}\}}$.  Therefore the left-hand side of Eq.\ \eqref{eq: equivalent expression of twist operator for even qubits} reduces to,

\begin{align}\label{}
     e^{-i\frac{\pi}{2}\frac{{(n-2k)}^2}{4}} &=  e^{-i\frac{\pi}{2}{(q-k)}^2} \nonumber \\
     &=e^{-i\frac{\pi}{2}{s}^2}
\end{align}\
where $s \in {\{1,2,.....q\}}$. Here \textit{q} and \textit{k} can be odd or even. If \textit{q} and \textit{k}, both are odd or even we get \textit{s} as even number and otherwise we get \textit{s} as odd number. Therefore we get two solution for the above case, i.e., 

\begin{equation}\label{eq:twist operator in s}
    e^{-i\frac{\pi}{2}\frac{{(n-2k)}^2}{4}} = \begin{cases}
    1 &  \text{when s is even, i.e., $(q-k)$ is even } \\
    -i &  \text{when s is odd, i.e., $(q-k)$ is odd} \\
\end{cases}.
\end{equation}

Now, the right hand side of the Eq.\ \eqref{eq: equivalent expression of twist operator for even qubits} takes a form,

\begin{align}\label{}
     e^{-i\frac{\pi}{4}}\Bigg[\frac{1 + (i)^{n+1+2k}}{\sqrt{2}}\Bigg] &=  e^{-i\frac{\pi}{4}}\Bigg[\frac{1 + (i)^{2(q+k)+1}}{\sqrt{2}}\Bigg]  
\end{align}
$q \in {\{1,2,.....\frac{n}{2}\}}$. Since $(2(q+k)+1)$ is always an odd number, we can split it into two forms, $(4l+1)$, when both \textit{q} and \textit{k} are even or odd (i.e., $(q-k)$ is even), and $(4l-1)$ otherwise (i.e., $(q-k)$ is odd) which results in,

\begin{equation}\label{}
    (i)^{n+1+2k} = \begin{cases}
    +i &  \text{when $2(q+k)+1) = 4l+1$ } \\
    -i &  \text{when $2(q+k)+1) = 4l-1 $} \\
\end{cases}.
\end{equation}
Therefore in this case we get two solution as well which are of form,

\begin{equation}\label{eq:twist operator in l}
    e^{-i\frac{\pi}{4}}\Bigg[\frac{1 + (i)^{n+1+2k}}{\sqrt{2}}\Bigg] = \begin{cases}
    1 &  \text{when  $(q-k)$ is even } \\
    -i &  \text{when  $(q-k)$ is odd} \\
\end{cases}.
\end{equation}
Since the solution of Eq.\ \eqref{eq:twist operator in s} and Eq.\ \eqref{eq:twist operator in l} are same, the relation mentioned in Eq.\ \eqref{eq:twist-jpi-integer} is proved.  

In the second part of proof we will show that Floquet operator show temporal periodicity. This can be done through repeated use of the Pauli group commutation relations. For any two elements \textit{g} and \textit{h} of a group G, a commutation relation is defined as,  

\begin{equation}\label{eq:commutator relations}
   [g,h] := g^{-1}h^{-1}g h
\end{equation}
We will first define the rotation part of the Floquet operator as,
\begin{equation}\label{eq:general rotation along y}
    e^{-i p J_y} = ( I \cos \frac{p}{2} - i \sigma_{y} \sin \frac{p}{2} )^{\otimes n} 
\end{equation}
for $p=\frac{\pi}{2}$, the above expression reduces to,

\begin{align}\label{eq:pi/2 rotation along y}
e^{-i \frac{\pi}{2} J_y} = ( I \cos \frac{\pi}{4} - i \sigma_{y} \sin \frac{\pi}{4} )^{\otimes n} := \gamma^{\otimes n}.
\end{align}
Therefore the commutation relation between $\gamma$ and $\sigma_z$ is given as 
\begin{align}\label{eq:gamma z commutation relation}
[\sigma_z,\gamma] &= \sigma_z^{-1}\gamma^{-1}\sigma_z\gamma \nonumber \\
&= i\sigma_y
\end{align}
This implies 
\begin{equation}\label{eq:gamma z relation}
  \sigma_z\gamma = - \gamma\sigma_x 
\end{equation}
Similarly,
\begin{equation}\label{eq:gamma x relation}
  \sigma_x\gamma = \gamma\sigma_z 
\end{equation}
We will use the above relations for our upcoming proofs. To show show the temporal periodicity, we will take the fourth power of the full unitary and compute its irreducible form. 

\begin{align}\label{eq:twist operator for even qubits with power 4}
    U_{\pi j}^4 &=  \left[\frac{e^{-i\frac{\pi}{4}}}{\sqrt{2}}\bigg( (I^{\otimes n} + i (i\sigma_z )^{\otimes n})\gamma^{\otimes n}\bigg) \right]^4 \nonumber \\ 
    &= \frac{e^{-i\pi}}{4}\Bigg[(I^{\otimes n} + i (i\sigma_z )^{\otimes n})\bigg( (I^{\otimes n} + i (i\sigma_x )^{\otimes n})\gamma^{\otimes n}\bigg)^3  \Bigg]\gamma^{\otimes n} \nonumber \\
    =& \frac{-e^{-i\pi}}{4}\Bigg[-4 (i\sigma_y )^{\otimes n})\Bigg] \nonumber \\
    =& - (i\sigma_y )^{\otimes n}
\end{align}
Using the above relation, the Floquet operator can be shown to exhibit the finite-time periodicity
\begin{equation}\label{}
    U_{\pi j}^8 = I \qquad \forall \, \, \text{integer } \, j.
\end{equation} 

\subsection{Half-integer spin}

In the case of half-integer spin and $\kappa = j\pi$ the general twist operator, Eq.\ \eqref{eq:general-twist-qubit}, happens to be equivalent to the unitary in the qubit picture given by Eq.\ \eqref{eq:twist-jpi-half-integer}.
In this case as well, we will show the equivalence relations by its actions on n-qubit state. Again twist operator will act in same way as Eq.\ \eqref{eq:twist-jpi-integer on hamming state}. The right hand side on the Eq.\ \eqref{eq:twist-jpi-half-integer} acts in the following way,
\begin{equation}\label{eq:equivalent twist-jpi-half-integer on hamming state}
    \begin{split}
    &\quad e^{-i\frac{\pi}{8}} \frac{1}{\sqrt{2}} \left[  e^{i\frac{\pi}{2}J_z} + e^{-i\frac{\pi}{2}J_z}  \right]\ket{D^{(k)}_n}  \\
    &= e^{-i\frac{\pi}{8}} \frac{1}{\sqrt{2}} \left[  e^{i(n-2k)\frac{\pi}{4}} + e^{-i(n-2k)\frac{\pi}{4}}  \right]\ket{D^{(k)}_n}
\end{split}
\end{equation}

Eq.\ \eqref{eq:twist-jpi-half-integer} and Eq.\ \eqref{eq:equivalent twist-jpi-half-integer on hamming state} implies,
\begin{equation}\label{eq: equivalent expression of twist operator for odd qubits}
      e^{-i\frac{\pi}{2}\frac{{(n-2k)}^2}{4}} = e^{-i\frac{\pi}{8}} \frac{1}{\sqrt{2}} \left[  e^{i(n-2k)\frac{\pi}{4}} + e^{-i(n-2k)\frac{\pi}{4}}  \right]   
\end{equation}
To prove the above relations, let's first see the left part of the Eq.\ \eqref{eq: equivalent expression of twist operator for odd qubits}. Since $n$ is odd and $2k$ is even, $(n-2k)$ will always be an odd and it will vary from $-n$ to $n$. Therefore, $(n-2k)^2$ will vary from 1 to $n^2$. Let's define $q$ such that,

\begin{equation}\label{eq: odd qubit as m for twist operator}
      n-2k:= 2q+1  
\end{equation}
where $q\in \{0,1,2,..,\frac{n-1}{2}\}$. Therefore,

\begin{align}\label{eq: odd qubit as m for twist operator mod 4}
     e^{-i\frac{\pi}{2}\frac{{(n-2k)}^2}{4}} &=  e^{-i\frac{\pi}{2}\frac{(2q+1)^2}{4}} \nonumber \\
     &=e^{-i\frac{\pi}{8}}e^{-i\frac{\pi q(q+1)}{2}}
\end{align}\
Here it can be seen that Eq.\ \eqref{eq: odd qubit as m for twist operator mod 4} takes a value $ e^{-i\frac{\pi}{8}}\{1,-1,-1,1\}$ for $q=\{0,1,2,3\}$ and it is same for  $q$ mod $4$.
Now considering the second part of the Eq.\ \eqref{eq: equivalent expression of twist operator for odd qubits}. It can be written as

\begin{equation}\label{eq: odd qubit as m for quivalent twist operator}
 \begin{split}
 &\quad e^{-i\frac{\pi}{8}} \frac{1}{\sqrt{2}} \left[  e^{i(n-2k)\frac{\pi}{4}} + e^{-i(n-2k)\frac{\pi}{4}}  \right] \\
 &= e^{-i\frac{\pi}{8}} \frac{1}{\sqrt{2}} \bigg[2 \text{ Cos} ((n-2k)\frac{\pi}{4})\bigg]
 \end{split}  
\end{equation}
where $(n-2k) \in \{-n, -(n+1),...,(n-1),n\}$. Since cos is an even function and $(n-2k)$ is odd of form ($2m+1$), the argument of cos function only varies from 1 to $n$. 

\begin{equation}\label{eq: odd qubit as m for equivalent twist operator mod 4}
\begin{split}
 &\quad e^{-i\frac{\pi}{8}} \frac{1}{\sqrt{2}} \left[  e^{i(n-2k)\frac{\pi}{4}} + e^{-i(n-2k)\frac{\pi}{4}}  \right] \\
 &= e^{-i\frac{\pi}{8}} \frac{1}{\sqrt{2}} \bigg[2 \text{ Cos} ((2q+1)\frac{\pi}{4})\bigg]
  \end{split} 
\end{equation}

Here also it can be seen that Eq. \eqref{eq: odd qubit as m for equivalent twist operator mod 4} takes a value $\{1,-1,-1,1\}$ for $q=\{0,1,2,3\}$ and it is same for  $q$ mod $4$. Therefore, the prove of Eq. \eqref{eq: equivalent expression of twist operator for odd qubits} is complete. 

 To show the temporal periodicity of the full Floquet operator, we will use the above expression of the twist operator given in  Eq.\ \eqref{eq:twist-jpi-half-integer} along with the rotation part of the unitary given by Eq. \eqref{eq:pi/2 rotation along y} and we took its sixth power. This gives us 

\begin{align}\label{eq:twist operator for odd qubits with power 6}
    U_{\pi j}^6 &=  \left[e^{-i\frac{\pi}{8}} \frac{1}{\sqrt{2}} \bigg(e^{i\frac{\pi}{2}J_z} + e^{-i\frac{\pi}{2}J_z} \bigg)\gamma^{\otimes n}\right]^6
\end{align}
We use the similar trick as in integer-spin case of pulling the rotation part of $U^6$ towards the end using group commutator relations. This leads us to an expression given by
\begin{align}\label{eq:final twist operator for odd qubits with power 6}
    U_{\pi j}^6 &= \frac{e^{-i\frac{3\pi}{4}}}{8}\Bigg[\bigg(e^{i\frac{\pi}{2}J_y} + e^{-i\frac{\pi}{2}J_y}\bigg)\bigg(e^{i\frac{\pi}{2}J_z} + e^{-i\frac{\pi}{2}J_z}\bigg)\Bigg]^3 (\gamma^{\otimes n})^6 \nonumber \\
    &= \frac{e^{-i\frac{3\pi}{4}}}{8 }A^3 (i\sigma_y)^{\otimes n}. 
\end{align}
where $A$ is composed of the terms in side the square bracket. Now we will compute the $A^2$ and we will show that $A^3$ just an $\mathbb I$ with some coefficient. For the next few steps, we have repeatedly used the following relations,

\begin{align}\label{eq:group commutator relation}
    e^{+ i\frac{\pi}{2}J_b }\ e^{-i\theta J_a }\ e^{-i\frac{\pi}{2}J_b } = \epsilon_{abc}\ e^{-i\theta J_c} \nonumber \\
    \text{where } (a,b,c) \in (x,y,z), 
\end{align}

\begin{align}\label{eq:special identity}
    \bigg(e^{\pm i\frac{\pi}{2}J_a }\ e^{\pm i\frac{\pi}{2}J_b  }\bigg)^3 = -\mathbb I \qquad \text{where } (a,b) \in (x,y,z), 
\end{align}

Let's write $A^2$ as
\begin{equation}\begin{split}\label{eq: A with power 2}
     A^2 &=
    (e^{-i\frac{\pi}{2}J_y}e^{- i\frac{\pi}{2}J_z})^2 \,+\,  2(e^{+i\frac{\pi}{2}J_y}e^{+i\frac{\pi}{2}J_x})\,+\, 2(e^{+i\frac{\pi}{2}J_x}e^{+i\frac{\pi}{2}J_z}) \\ 
    & \,+\,  2(e^{+i\frac{\pi}{2}J_x}e^{-i\frac{\pi}{2}J_z})\,+\, (e^{+i\frac{\pi}{2}J_y}e^{-i\frac{\pi}{2}J_z})^2 \,+\, (e^{-i\frac{\pi}{2}J_y}e^{-i\frac{\pi}{2}J_x})\\
    &  \,+\, 2(e^{-i\frac{\pi}{2}J_x}e^{-i\frac{\pi}{2}J_z})^2\,+\, (e^{-i\frac{\pi}{2}J_x}e^{+i\frac{\pi}{2}J_z})\,+\, (e^{-i\frac{\pi}{2}J_y}e^{+i\frac{\pi}{2}J_z})^2\\
    & \,+\, (e^{-i\frac{\pi}{2}J_y}e^{+i\frac{\pi}{2}J_x}) \,+\, (e^{-i\frac{\pi}{2}J_y}e^{- i\frac{\pi}{2}J_x})^2 \,+\, (e^{-i\frac{\pi}{2}J_y}e^{-i\frac{\pi}{2}J_z})^2. 
\end{split}\end{equation}
Multiplying the above expression with $A$, we get,
\begin{align}\label{eq:A as identity with power 3}
    A^3 = 8\,\mathbb I.
\end{align}
Using Eq. \eqref{eq:twist operator for odd qubits with power 6} and Eq. \eqref{eq:A as identity with power 3} we get,

\begin{align}\label{eq:floquet operator for odd qubits as identity with power 6}
    U_{\pi j}^6 &= e^{-i\frac{3\pi}{4}} (i\sigma_y)^{\otimes n}. 
\end{align}

Finally we get,
\begin{align}\label{eq:floquet operator for odd qubits as identity with power 12}
    U_{\pi j}^{12} &= e^{-i\frac{\pi}{2}}\mathbb I. 
\end{align}

\section{Stability}

In the main section we showed that for certain values of the chaoticity parameter $\kappa$ (see Table.\ \ref{summary_table}), the quantum kicked top unitary experiences a state-independent temporal periodicity and therefore does not reflect the classical dynamics, chaotic or otherwise.  The next natural step is to consider what happens as the twist strength is slightly detuned from the exact recurrence values found above.  Here we discuss the stability of these special values $\Tilde{\kappa}$ via perturbations of the form $\kappa = \Tilde{\kappa} + \delta$, where $\delta$ is some small deviation incommensurate to $\pi$.  The Floquet operator becomes
\begin{eqnarray}\label{eq:floquet-unitary_perturbed}
\begin{split}
        U^N_{\Tilde{\kappa} + \delta} &= \Bigg[ \exp\big(-i\frac{\Tilde{\kappa} + \delta }{2j}J_{z}^2\big) \exp\big(-i\frac{\pi }{2}J_{y}\big) \Bigg]^N \\
     &= \Bigg[ \exp\big( -i\frac{\delta }{2j}J_{z}^2 \big) \cdot U_{\Tilde{\kappa}} \Bigg]^N 
\end{split}
\end{eqnarray}
where $N$ is the orbit length of the periodicity at $\Tilde{\kappa}$ summarized in Table \ref{summary_table}.

\begin{figure}[h!]
  \centering
  \includegraphics[width=0.8\columnwidth]{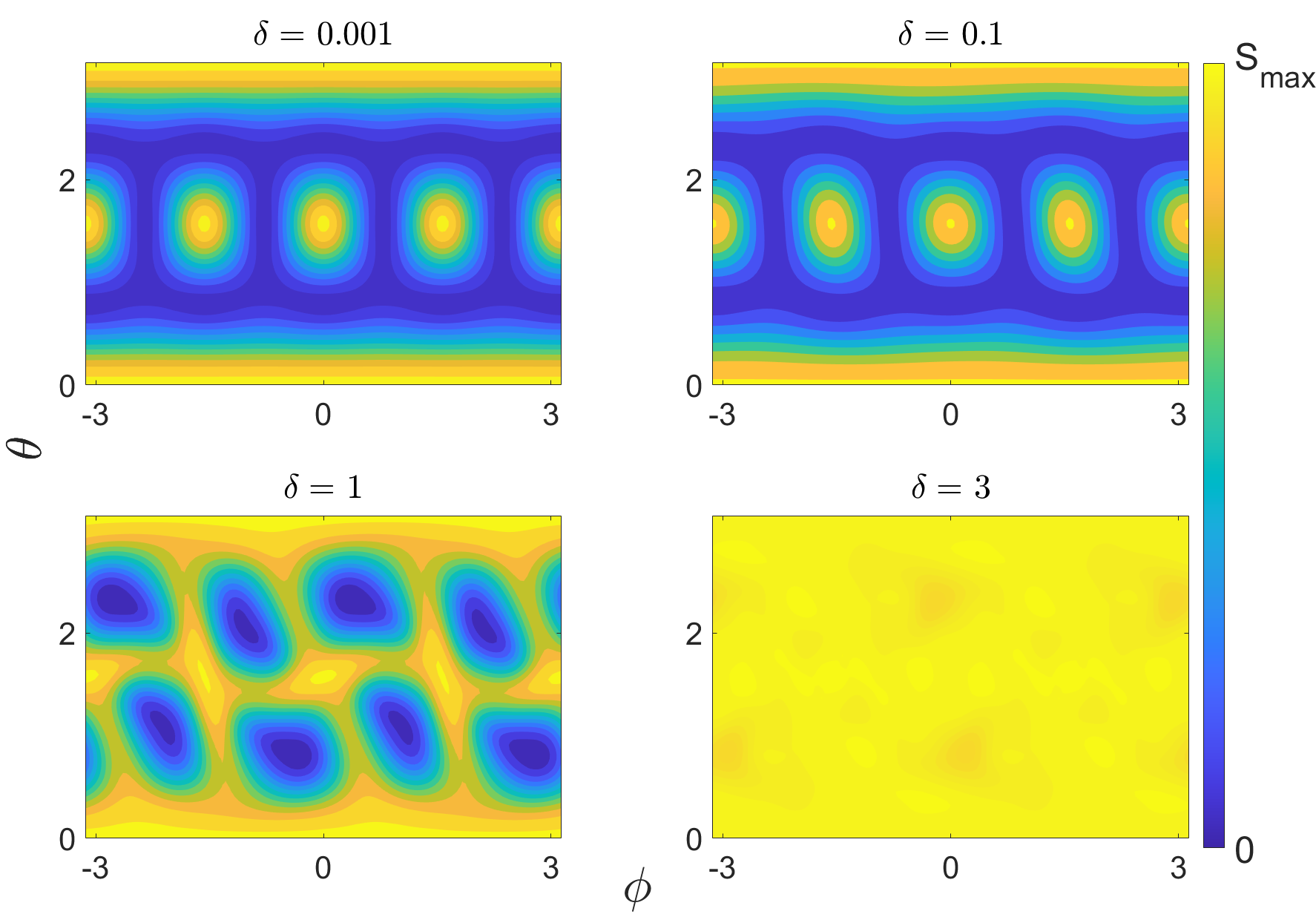}
  \caption{Average entanglement entropy with $j=15.5$ and $\kappa = \pi j + \delta$, calculated on a grid of 70 x 140 initial spin coherent states. Each initial state is time-averaged over 10 applications of $U^{12}_{\pi j + \delta}$ to see the cumulative effect of the error $\delta$. $S_{\text{max}} \in \{7\times10^{-11}, 2\times10^{-3}, 0.6097, 0.6868\}$ is the maximum entropy achieved for $\delta \in \{0.001, 0.1, 1, 3\}$, respectively.} 
  \label{fig:perturbed_entropy_landscape_half_int}
\end{figure}

We proceed by studying the entanglement entropy $S$ of one of the reduced qubits.  As spin coherent states are product states in the qubit picture, when there is no perturbation ($\delta = 0$) the entropy of the state $U^N_{\Tilde{\kappa}}\ket{\theta,\phi}$ must vanish.  Hence as $\delta$ slowly increases so too must the entropy, reflecting the change in dynamics from finite recurrence to standard kicked top evolution.  Fig.\ \ref{fig:perturbed_entropy_landscape_half_int} shows the entropy for $j=15.5$, $\kappa=\pi j + \delta$ with different $\delta$ values, averaged over the application of $U^{N}_{\kappa+\delta}$ 10 times.  That is, 
\begin{equation}
  \frac{1}{10} \sum_{n=1}^{10} S( U^{12n}_{\pi j + \delta} \ket{\theta,\phi} ).
\end{equation}
This coarse-grained evolution and averaging is done to capture the cumulative effect of the perturbation $\delta$ on the ability of the orbit to return a spin coherent state after $N$ kicks (where in this example $N=12$ because the spin is half-integer and $\kappa=\pi j$).

There is a clear change in the average dynamics as the perturbation increases, with the relatively strong value of $\delta=3$ showing an entropy plot that matches the phase space structure of the highly chaotic classical kicked top associated to $\kappa=15.5\pi$.  This gradual change is attributed to the shearing \cite{kitagawa_1993_squeezed} from the intermediate squeezing operator $e^{-i\frac{\delta}{2j}J_z^2}$ \eqref{eq:floquet-unitary_perturbed} on each superposed spin coherent state arising from $U_{\Tilde{\kappa}}$.  This intermediate shearing delocalizes the constituent spin coherent states over phase space, which then inhibits their recombination effect typical of Fig.\ \ref{fig:integer_j_Husimi_evolution_jpi}.

\begin{figure}[h]
  \centering
  \includegraphics[width=0.7\columnwidth,trim={2.5cm 0 0 0},clip]{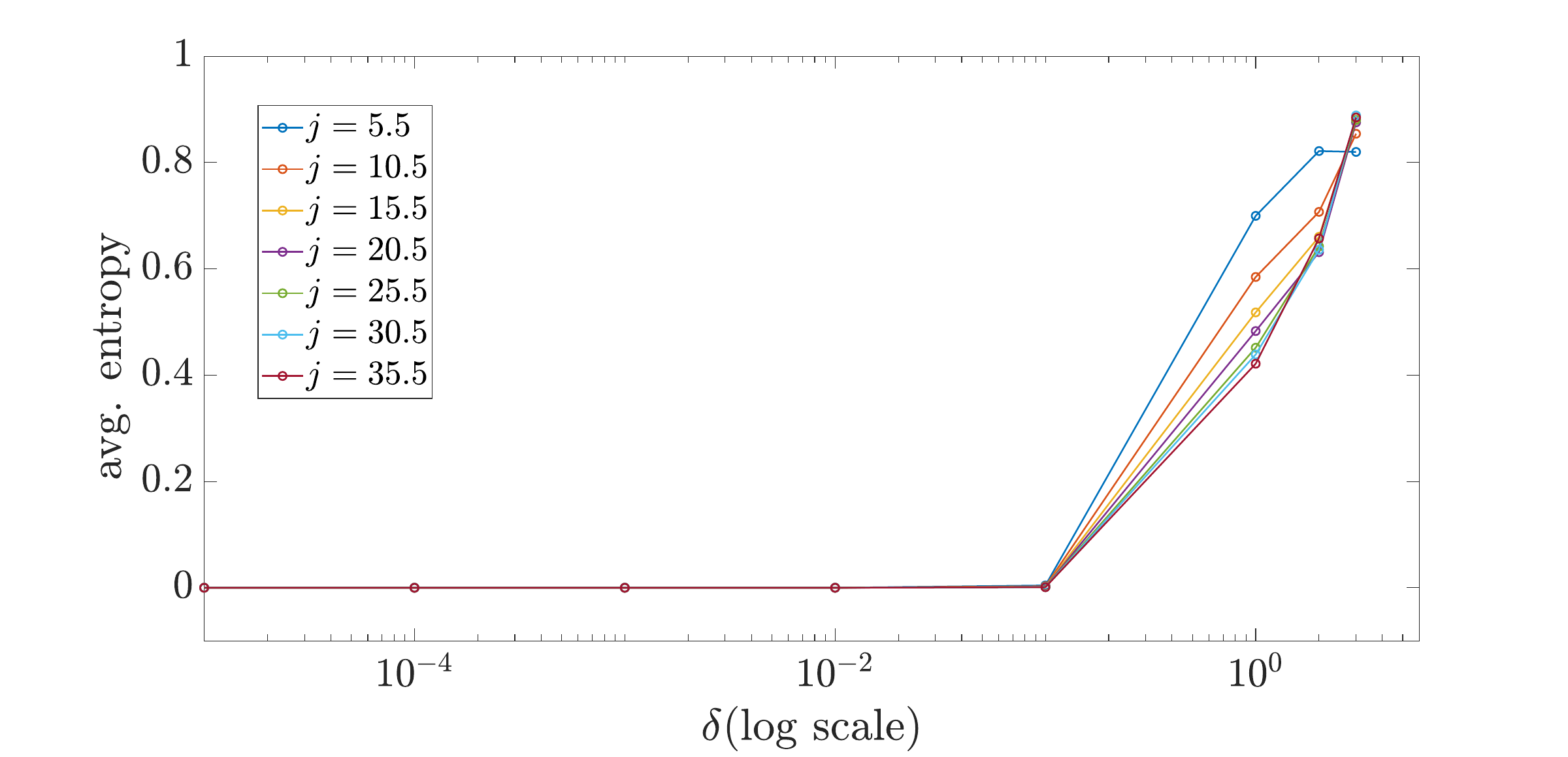}
  \caption{Mean value of time average entanglement entropy for different spin ($j$) and $\kappa = \pi j + \delta$, calculated on a grid of 70 x 140 initial spin coherent states. Each initial state is time-averaged over 10 applications of $U^{12}_{\pi j + \delta}$.}
 \label{fig:average_entropy_15.5}
\end{figure}

We have also found that the rate of the transition between periodic dynamics and standard dynamics (i.e.\ as $\delta$ increases) largely depends only $\Tilde{\kappa}$ and not directly on $j$.  Fig.\ \ref{fig:average_entropy_15.5} shows the mean value of time-averaged entanglement entropy over the entire phase space with  70 x 140 initial spin coherent states. It can be observed that for all values spin $j$, system becomes unstable at $\delta=0.01$. The relation between $\delta$ and instability of these special parameters can be explored in future works.   

\end{document}